\documentclass{article}

\PassOptionsToPackage{sort&compress,numbers,super}{natbib}
\usepackage[preprint]{neurips_2023}
\usepackage[utf8]{inputenc} 
\usepackage[T1]{fontenc}    
\usepackage{hyperref}       
\usepackage{url}            
\usepackage{booktabs}       
\usepackage{amsfonts}       
\usepackage{nicefrac}       
\usepackage{microtype}      
\usepackage{xcolor}         
\usepackage{lmodern}
\usepackage[pdftex]{graphicx}

\usepackage{adjustbox}
\usepackage{bbding}
\usepackage{array,booktabs}
\usepackage{enumitem,amssymb}
\newlist{todolist}{itemize}{2}
\setlist[todolist]{label=$\square$}
\usepackage{caption}
\usepackage{tabularx}
\usepackage{comment}

\title{ Extracting human interpretable structure-property relationships in chemistry using XAI and large language models}

\author{
  Geemi P.~Wellawatte \qquad Philippe Schwaller\\
    Laboratory of Artificial Chemical Intelligence \\
  Institute of Chemical Sciences and Engineering \\
  Ecole Polytechnique Fédérale de Lausanne (EPFL)\\
  Lausanne, Switzerland\\
  \texttt{\{geemi.wellawatte,philippe.schwaller\}@epfl.ch} \\
}

\begin{document}

\maketitle

\begin{abstract}
Explainable Artificial Intelligence (XAI) is an emerging field in AI that aims to address the opaque nature of machine learning models. Furthermore, it has been shown that XAI can be used to extract input-output relationships, making them a useful tool in chemistry to understand structure-property relationships. However, one of the main limitations of XAI methods is that they are developed for technically oriented users. We propose the XpertAI framework that integrates XAI methods with large language models (LLMs) accessing scientific literature to generate accessible natural language explanations of raw chemical data automatically. We conducted 5 case studies to evaluate the performance of XpertAI. Our results show that XpertAI combines the strengths of LLMs and XAI tools in generating specific, scientific, and interpretable explanations. 

\end{abstract}

\section{Introduction}

Understanding structure-property relationships has been a long-standing challenge in chemistry\cite{wiener1947structural,seybold1987molecular,mihalic1992graph}
Seybold et al.\cite{seybold1987molecular} highlight a fundamental concept in chemistry: the ``properties and behaviors of molecules follow from their structures''. 
Therefore, elucidating these relationships facilitates the manipulation of molecules to achieve desired properties.  
Machine learning (ML) is a routinely used tool to complement human expertise, which solves complex tasks in chemistry by modeling structure-property relationships.\cite{ren2023alphafold,kim2021visual,gawehn2016deep,lysenko2018integrative,Jaiswal2021} 
While ML has been proven to be successful in solving such tasks in chemistry,\cite{deringer,faber,Gupta2021,Duch2007, Dara123,Webb,Gomes} experimental chemists often harbor skepticism toward predictions generated by such models, primarily due to the inherent opacity of these models. 
In essence, these ML models usually do not provide a rationale as to why a certain prediction was made. 
EXplainable Artificial Intelligence (XAI) is a new branch of AI that is rapidly growing and aims to explain the opacity nature of ML models.
Therefore, developing XAI tools for chemistry is critical for increasing trust in ML models and expanding the possibilities of experimental and computational chemistry. 

Justifications, explanations, and interpretability are three terms associated with XAI.\cite{murdoch2019definitions,schwalbe2023comprehensive,wellawatte2023} While a justification simply provides evidence for a prediction\cite{miller2019explanation}, an explanation describes the rationale for the prediction.\cite{biran2017explanation}  
However, the true potency of XAI lies in its interpretability, which concerns the extent to which a human can comprehend the provided explanation.\cite{Murdoch2019InterpretableML} 
In a recent survey, Cambria et al.\cite{CAMBRIA2023103111} emphasized that there is a pressing need to refine the presentation of explanations. 
This means that although XAI addresses the opacity of ML models, they are not user-friendly for non-domain experts or non-technical users. 
Therefore, there is growing interest in incorporating natural language (NL) with XAI to produce more accessible explanations.\cite{CAMBRIA2023103111,mariotti2020towards} 
Furthermore, it's worth noting that existing XAI methods often lack the flexibility to address specific user queries -- can usually answer only one specific question, impeding their adaptability.\cite{kanehira2019learning,Sheridan2019,russell2019efficient,Wellawatte2022}

To meet the demand for creating intelligent, adaptable, and user-friendly XAI tools for chemistry, we introduce a Python package named ``XpertAI''. 
Our tool combines XAI methods with large language models (LLMs) to extract structure-property relationships from raw data. 
LLMs are generative models which can predict an output sequence given an input sequence. LLMs can be made into powerful agents that query databases, scrape and summarize literature, interpret, and generate text in NL.\cite{white2023future} 
However, LLMs in isolation are limited in addressing domain-specific problems within the field of chemistry. 
For example, in the work by Hendricks et al.,\cite{hendricks2021generating}, the authors provide a promising workflow for generating NLEs in image classification tasks. 
However, such approaches are not readily adaptable in chemistry applications due to a number of limitations, including the molecular representation issue commonly associated with deep learning in chemistry.
To circumvent such challenges at the intersection of chemistry and LLMs, Jablonka et al.~\cite{jablonka2023gpt} demonstrated that finetuning LLMs could provide a solution to this.
Additionally, Bran et al.~\cite{bran2023chemcrow} showed LLMs can be enhanced to tackle tasks such as organic synthesis, drug discovery, and materials design.
Previously, it has been suggested that ``black-box modeling first, followed by XAI'' as a means to establish structure-property relationships without compromising accuracy or interpretability.\cite{wellawatte2023} 
Seshadri et al.\cite{aditi2022} showed that LLMs combined with XAI can generate human interpretable explanations. 
Unlike our approach, this work used LLMs only to summarize the findings from the XAI analysis in natural language. 
Following a similar direction, we demonstrate that LLMs combined with XAI tools and literature evidence, play a powerful role in generating both interpretable and scientifically accurate explanations.

\begin{figure}[!ht]
  \centering
  \includegraphics[width=1\linewidth]{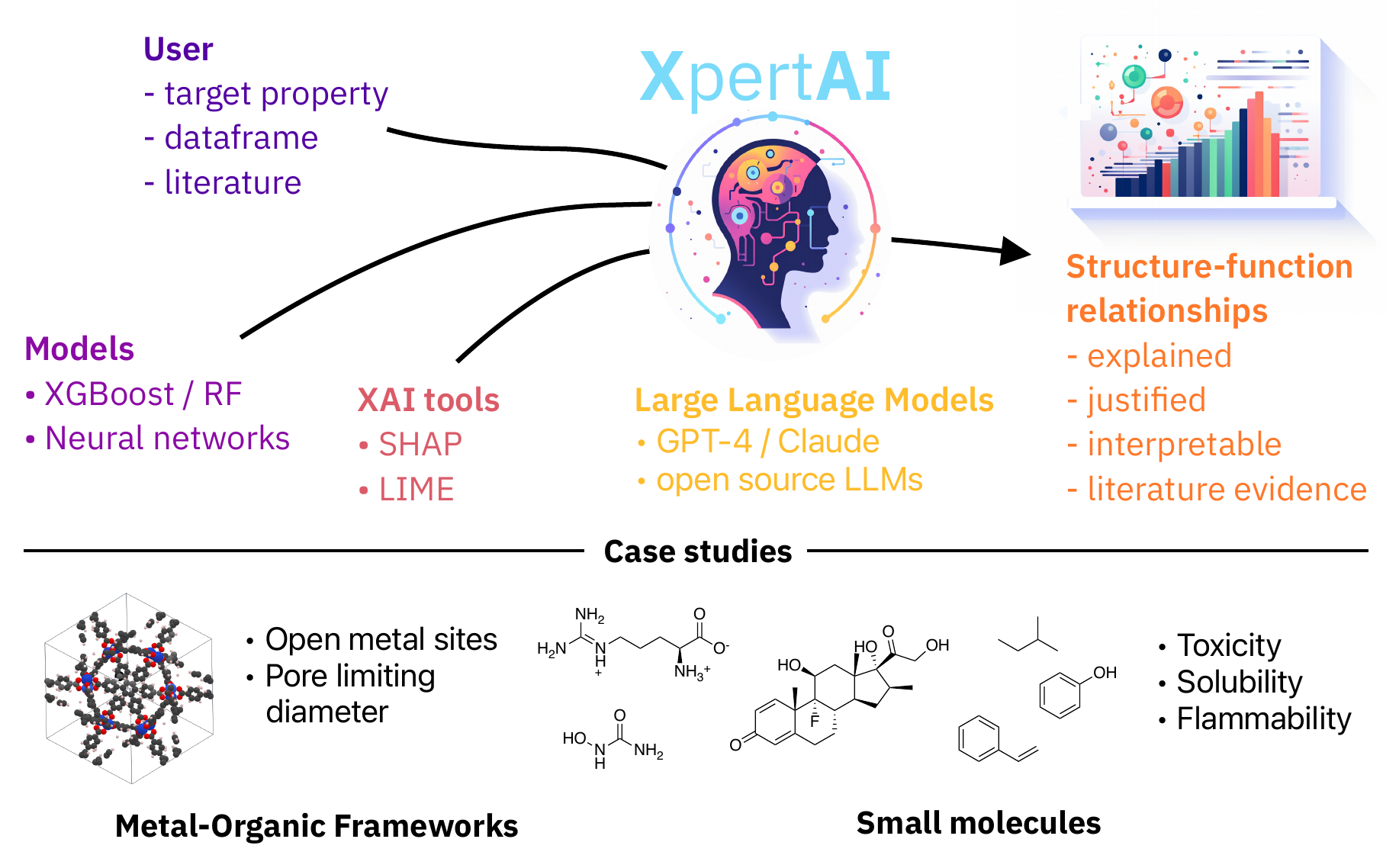}
  \caption{Overview of XpertAI. The proposed tool combines XAI with LLMs to uncover human interpretable structure-property relationships from raw data.}

  \label{overview}
\end{figure}

\begin{figure}
  \centering
  \includegraphics[width=0.9\linewidth]{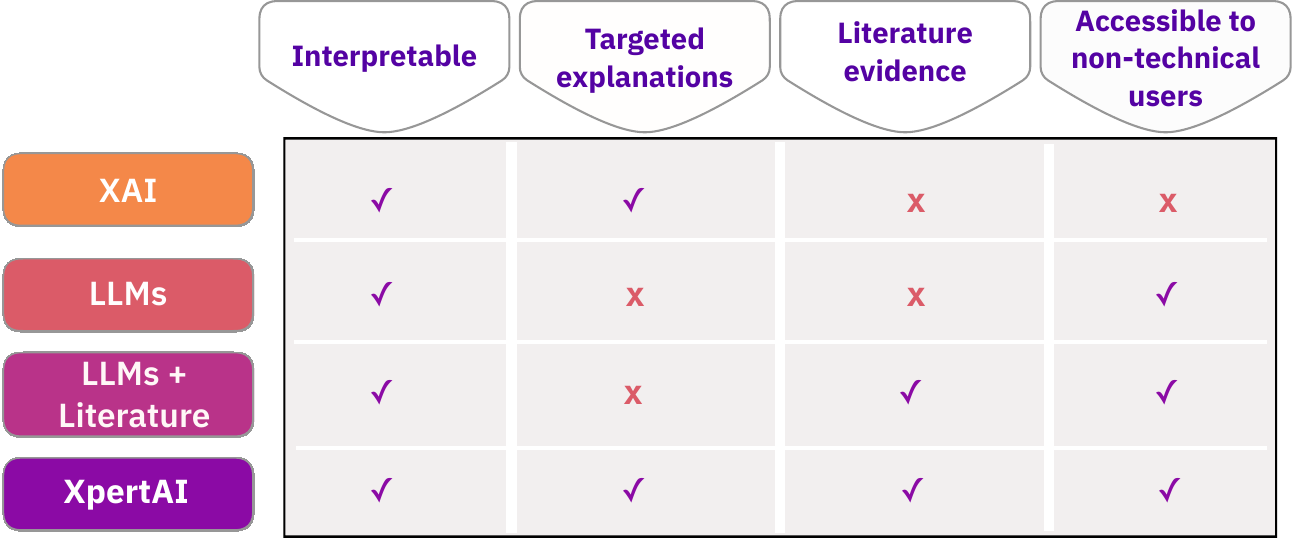}
  \caption{Attributes of XpertAI explanations in comparison to baseline methods.}

  \label{tab}
\end{figure}

XpertAI aims to establish connections between black-box models, XAI tools, and literature through LLMs to uncover relationships between molecular features and target properties. 
As shown in Figure \ref{overview}, XpertAI employs XAI methods to identify crucial features within the dataset and subsequently drawing on scientific evidence to articulate structure-property connections based on these features. 
One of the key advantages of XpertAI is its ability to deliver precise natural language explanations (NLEs) tailored to specific datasets, as opposed to providing generalized explanations drawn from the broader literature. 
As illustrated in Figure \ref{tab} XpertAI combines the strengths of XAI and LLMs in terms of specificity, interpretability, accessibility, and scientific nature of the explanations. 
To the best of our knowledge, currently there is no such tool in chemistry, which extracts NL structure-property relationships from user-given raw data. 
Furthermore, our application is generalizable to any domain that requires extracting input-output relationships as NLEs. 

\section{Method}
We begin the workflow by training an ML model using the initial raw data. This model serves as a surrogate for mapping input to output. 
The initial dataframe includes featurized molecular structures and target labels for training.
Note that these features must be human-interpretable (e.g., molecular descriptors/properties or MACCS keys). 
We currently employ gradient-boosting decision trees with the XGBoost framework, utilizing the Scikit-learn API for regression and classification tasks.\cite{xgboost,scikit} 
Once the model is trained, users can select from SHAP\cite{lundberg2017unified}, LIME\cite{ribeiro}, or both to estimate the ``most impactful'' features. 
SHAP and LIME are commonly used to generate local explanations.
Therefore, we compute mean SHAP values and Z-scores for LIME explanations to extract globally impactful features. 
For the LIME analysis, we only use a sample of the initial dataset due to time and resource constraints.
The default sample size is either 500 or the entire dataset if its length is less than 500. 
After identifying impactful features, we draw knowledge from the literature to elucidate physicochemical relationships between these features and the target property. 
As seen in the overview of our proposed workflow (Figure \ref{overview}) LLMs are used to unite the backend modules generating human interpretable explanations. 
We leverage on LangChain python package (\url{https://github.com/langchain-ai/langchain}), OpenAI's GPT-4\cite{gpt4} language model, and Chroma vector database (\url{https://github.com/chroma-core/chroma}) to query multiple literature publications and generate intelligent explanations. 
Users have the flexibility to upload a literature dataset or scrape \url{arxiv.org} to gather relevant literature information. 
The latter is enabled via the arXiv python API. 

Once the literature is collected, we select the most related literature excerpts using maximal marginal relevance search (MMR) to validate XAI observations. 
Following a similar approach as the ``StuffDocumentsChain'' in LangChain, we stuff these excerpts to a specialized prompt to generate the complete explanation. 
We utilize the chain-of-thought prompting approach\cite{wei2022chain} to create the NLEs -- we provide a series of intermediate steps and examples within the prompt to improve the output NLE's interpretability. 
XpertAI also generates and adds citations in the final NLEs to improve the accountability of the explanations. 
We would like to highlight that, in addition to the NLEs, XpertAI also provides the surrogate model's evaluation plot (error plot) and XAI analysis plots for the users.
To streamline this complete workflow, we have deployed a Streamlit  App (\url{https://xpert-ai.streamlit.app/}) that can be used with an OpenAI API key. 
The GitHub repository of our work can be found at: \url{https://github.com/geemi725/XpertAI}. 

\section{Results}
We used the XpertAI tool to suggest structure-property relationships for five case studies in chemistry; 1) the presence of open metal sites in metal-organic frameworks (classification), 2) pore limiting diameter in metal-organic frameworks (regression), 3) toxicity of small molecules (classification), 4) solubility of small molecules (regression), and  5) upper flammability limit of organic molecules (regression). 
Please note that we used the SHAP method as the chosen XAI method, and the default hyperparameters to generate NLEs in the following case studies. We chose the SHAP method due to its consistency in generating global explanations in comparison to LIME.
Complete NLEs and SHAP plots for each case study from XpertAI are provided in Appendices A and D respectively. 

\textbf{Case studies 1 and 2: Structure-property relationships in Metal-organic frameworks (MOFs).} 
MOFs, a hybrid class of materials in chemistry, consist of metal nodes connected by organic linkers.\cite{hall2019structure} Their porous nature lends them versatile properties such as gas separation and storage\cite{Ding2019,schoedel2016role,wang2016imidazolium}, catalysis\cite{lee2009metal,yang2019catalysis}, and drug delivery\cite{horcajada2008flexible,horcajada2012metal}. 
Understanding MOF structure-property relationships is crucial for optimizing their design in specific applications.
Open metal sites, characterized by coordinative unsaturation, find valuable use in catalysis.\cite{hall2019structure}
Additionally, the pore-limiting diameter is a key feature for screening them in selective gas capture applications.\cite{hung2021transport} 
However, the precise relationship between MOF atomic structure and open metal sites or pore-limiting diameter remains incompletely understood.

In case study 1, we sampled 3734 MOFs from the CoRE MOF 2019 database\cite{chung2019advances}.  
These crystal structures obtained as CIF files were then featurized using the CrystalFeatures tool\cite{tawfik2022naturally}.
Generated features are meaningful descriptors encompassing atomic and crystal characteristics, geometry features, and one-shot ab initio descriptors.
Next, we uploaded the featurized inputs along with the binary target labels (presence/absence of open metal sites) as a CSV dataframe along with a literature dataset containing 19 publications to our XpertAI Streamlit App. 
The list of publications can be found in our GitHub repository.
The generated NLE from XpertAI explains how a) metals fraction, b) density of solid, c) average cationic radius correlate with the presence of open metal sites. 
This explanation aligns with the findings of Hall et al.\cite{hall2019structure}, where the authors identify metal identity and oxidation state, defect density, and site proximity to be impactful structural components. 
Note that we did not include this review paper\cite{hall2019structure} in the literature dataset uploaded to XpertAI to avoid data tampering. 
Although the input raw dataset did not include features such as the fraction of electrons in d orbitals, the average ionization energy, and the average electron affinity, XpertAI suggests these features to be correlated with the open metal sites based on literature data.

Following a similar approach in case study 2, we used the same MOF dataset but with pore-limiting diameters as the label. Unlike case study 1, this is a regression-type problem. 
According to the XpertAI explanation, volume per atom, symmetry function G, and unoccupied energy levels at the conduction band influence the pore-limiting diameter. 
According to XpertAI, ``The Volume Per Atom can affect the Pore-limiting diameter in MOFs as it determines the size of the atoms in the MOF structure. Larger atoms may result in smaller pore
sizes (Haldoupis, Nair, \& Sholl, 2010)\cite{haldoupis2010efficient}. The symmetry function G is a measure of the symmetry of the MOF structure, which can also influence the Pore-limiting diameter. MOFs with higher symmetry may have more uniform pore sizes (Yang et al., 2022).\cite{yang2022analyzing}''
Please see Appendix A for the complete textual explanation.

\textbf{Case study 3: Small molecule toxicity.} 
Toxicity prediction of small molecules is a benchmark task in chemistry, particularly in drug discovery.\cite{huang2016tox21challenge,durant2002reoptimization}
Despite the extensive research in this area, a precise understanding of the relationship between molecular structure and toxicity remains elusive. 
In this case study, we used 1478 molecules from the Tox21 database\cite{huang2016tox21challenge} where binary labels indicate toxicity (a classification task). 
Then we featurized the input molecules in SMILES format using  MACCS descriptors\cite{durant2002reoptimization} implemented in the RDKit package\cite{landrum2016rdkit}.
These descriptors are human-interpretable binary features containing 167 yes/no questions regarding molecular structure. 
Next, we provided a curated literature dataset with 17 related publications.
Our XpertAI explanation summarizes: ``The features identified by the XAI analysis that affect the toxicity of small molecules include the presence of a heteroatom bonded to three Oxygen atoms, the presence of an aminomethyl group, and the presence of a Tertiary Amine. Additional features that may be correlated with the toxicity of small molecules include the solubility of the molecule, the presence of aromatic rings, and the presence of nucleophilic N, O, and aryl containing substructures.''
XpertAI suggests that these features alter toxicity as they affect the reactivity of the molecules and their ability to form reactive species, which can cause cellular damage and disrupt cellular processes.
This explanation aligns with the findings in work by Meanwell\citenum{meanwell2015influence} and, Limban et al.\cite{limban2018use} which state that aromatic amines and nitro groups are associated with increasing molecular toxicity. 
Please note that these references were not included in the literature dataset uploaded to XpertAI. 
The complete XpertAI NLE can be found in Appendix A.
Additionally, XpertAI suggests how to reduce the toxicity of a molecule -- the toxicity of a molecule can be reduced by removing the aminomethyl group or the tertiary amine because these increase the reactivity of the molecule, potentially leading to increased toxicity. 

\textbf{Case study 4: Small molecule solubility.} Aqueous solubility of small molecules is a critical property in drug discovery as solubility determines the interaction of the drug in a biological environment.\cite{peterson2001comparison} 
To explain the relationship between the molecular structure and its solubility, we used a sample dataset with 9982 molecules from the AqSolDB\cite{sorkun2019aqsoldb} dataset for training. 
Once again, we used MACCS descriptors to convert the molecules into a binary vector. 
We uploaded a literature dataset with 18 related publications. References to these can be found in our GitHub repository. 
XpertAI explains the structure-solubility relationship as follows. 
``The solubility of small molecules is a complex property that is influenced by a variety of factors. The XAI analysis has identified three key features that significantly impact the solubility of small molecules: the presence of an atom at an aromatic/non-aromatic boundary, the presence of two heteroatoms bonded to each other, and the presence of an atom with three heteroatom neighbors. Presence of an atom at an aromatic/non-aromatic boundary: This feature refers to the location of an atom at the boundary between an aromatic ring and a non-aromatic part of the molecule. Aromatic rings are typically hydrophobic, while non-aromatic parts of the molecule can be either hydrophilic or hydrophobic. Therefore, an atom at this boundary can influence the overall solubility of the molecule by affecting its polarity and ability to form hydrogen bonds with water molecules (Lee at al., 2023\cite{lee2023multi}).'' 
This excerpt demonstrates that XpertAI provides self-explanatory explanations with references to relevant literature. 

\textbf{Case study 5: Compound flammability.}The upper flammability limit (UFL) of an organic compound is an important characteristic that determines the hazardous nature of the compound.\cite{yuan2019developing} This is an interesting topic of study in both experimental and computational studies.\cite{mannan2012lees,vidal2004review,gharagheizi2009prediction,pan2009prediction}
We used the UFL dataset used by Yuan et al.\cite{yang2019catalysis}, which was extracted from reference Crowl and Louvar\cite{crowl2001chemical}. This dataset only contained 79 organic compounds. 
We used the same quantum and non-quantum molecular descriptor set used in reference \citenum{yuan2019developing} to featurize the molecules. 
After uploading the initial dataset and 11 curated publications, we obtained the following explanation. 
``The Structural Information Content Index (Neighborhood Symmetry of Order Zero) and Information Content Index (Neighborhood Symmetry of Order Zero) are measures of the complexity and symmetry of the molecule's structure. Molecules with higher complexity and less symmetry may have higher upper flammability limits due to the increased number of potential reaction pathways (Yuan et al., 2019)\cite{yuan2019}. 
The highest occupied molecular orbital (HOMO) represents the energy level of the highest energy electron in the molecule. Molecules with higher HOMO energy levels may have higher upper flammability limits because they have more energy available for combustion reactions (Yuan et al., 2019)\cite{yuan2019}. The Dipole Moment is a measure of the molecule's polarity. Molecules with higher dipole moments may have lower upper flammability limits because the increased polarity can inhibit the dispersion of the molecule in the air, reducing the likelihood of combustion (Yuan et al., 2019).\cite{yuan2019}'' 
Based on this explanation, we can conclude that XpertAI provides complete, textual explanations of molecular structure-function relationships extracted from raw data. As these explanations are equipped with literature references, users can evaluate the accuracy of the explanations easily. 

\section{Evaluation}
\begin{figure}
  \centering
  \includegraphics[width=1\linewidth]{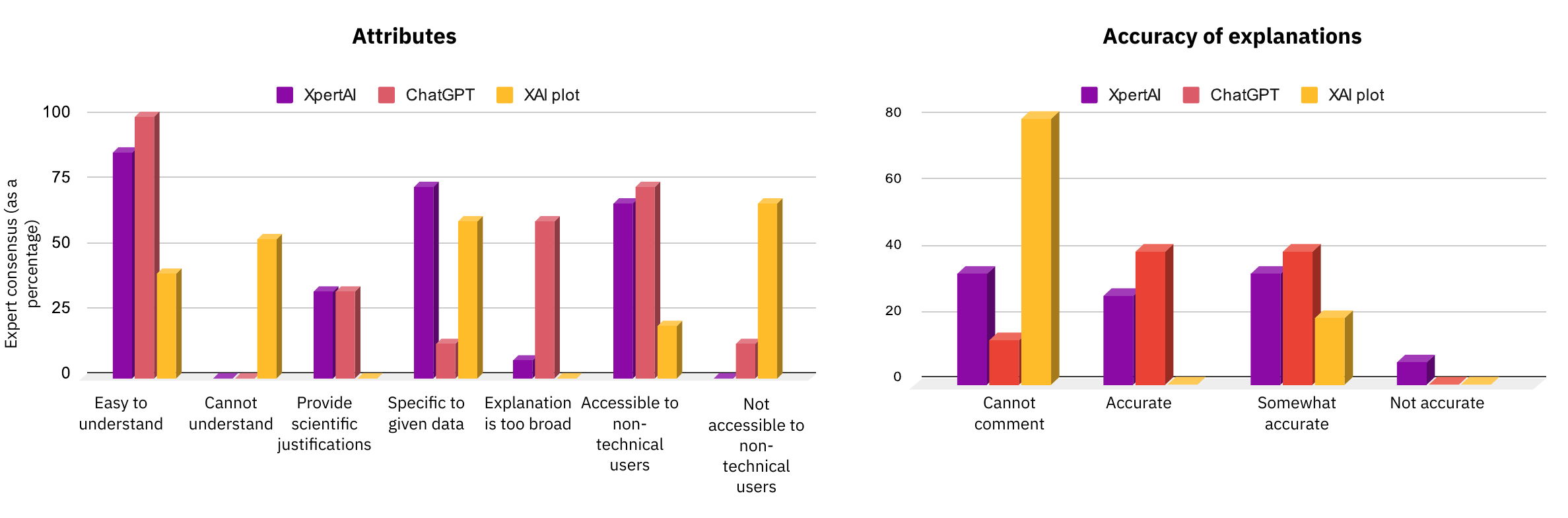}
  \caption{Human expert consensus. 3 human experts were asked to evaluate explanations from XpertAI, ChatGPT, and SHAP plots for each case study based on accuracy and interpretability. Expert consensus is given as a percentage. }

  \label{eval}
\end{figure}
Firstly, to evaluate the explanations for the listed case studies, we compared three different explanations from; 1) XpertAI 2) ChatGPT (GPT-4), and 3) Graphical plots from the XAI analysis (See Appendix D for SHAP plots). 
We selected these three setups to evaluate if combining LLMs with XAI can harness the advantages of both approaches rather than using one alone. 
We asked three expert chemists to score 15 explanations in total (5 tasks $\times$ 3 explanations). The experts were given a scorecard (given in Appendix C) to evaluate each answer based on accuracy and interpretability. 
As seen in Figure \ref{eval}, evaluators associated XpertAI NLEs with being specific to a given dataset/task, providing scientific explanations, and having high accessibility to non-technical experts. 
However, LLMs by themselves (ChatGPT) were associated with a lack of specificity to the given data, and XAI with a lack of interpretability.
As summarized in Figure \ref{tab}, the evaluations conclude that XpertAI effectively leverages the advantages of both ChatGPT and XAI to provide a complete explanation.
These results validate the accomplishment of our goal to extract accessible and interpretable structure-property relationships in chemistry from raw data.

Based on the expert scores, we note that XpertAI's explanations are considered ``accurate'' or ``somewhat accurate''. 
Although ChatGPT scored higher in terms of accuracy, we argue this is because human experts excel at validating generalized scientific hypotheses. 
However, XpertAI's explanations are specific to relationships established from provided datasets which may be different from familiar concepts in chemistry.
Therefore, we conclude that XpertAI shows promise in generating interpretable and accurate natural language descriptions of structure-function relationships.
Furthermore, we noted that the explanations remained consistent, indicating less probability of hallucinations over multiple generations. 
However, it should be emphasized that the accuracy and specificity of the generated explanations are dependent on the quality of the literature dataset provided.
To maintain the quality of explanations, we only used specifically curated publications in case studies 1-5. 
The lists of references used can be found in our GitHub repository.

Next, we asked Claude AI assistant (\url{https://claude.ai/}) by Anthropic \url{https://www.anthropic.com/} to compare the two explanations from XpertAI and ChatGPT based on accuracy and interpretability. The responses from Claude are given in Appendix B. 
Based on the responses, Claude rates the explanations from XpertAI higher than ChatGPT's for 4/5 tasks. We list a few reasons for Claude's higher ratings for XpertAI below. 
Note that we anonymized the two explanations during the scoring; Explanation A is by XpertAI and Explanation B is by ChatGPT.
\begin{itemize} 
    \item The language and terminology used in Explanation A is more accessible and easier to understand
    for a non-expert compared to Explanation B.
    \item Explanation A provides more details on the specific features that impact the target property, including both the top features from XAI analysis and other correlated features (from literature). In contrast, Explanation B only broadly refers to a few features.
    \item Explanation A provides mechanistic explanations and states how the target property can be altered by changing each feature in the desired direction. Explanation B lacks this level of mechanistic detail and only generally mentions that the features can be changed, but not how.
\end{itemize}
However, Claude rates XpertAI's explanation in case study 5 less than ChatGPT's. 
A similar observation was made from the expert evaluations, where 1 out of 3 evaluators indicated the XpertAI's NLE as not accurate.
According to Claude ``Explanation B demonstrates strong domain knowledge by linking features to concepts like vapor pressure, steric hindrance, and flame retardants. Explanation A does not connect to these chemistry concepts.'' 
We hypothesize this is due to the underperformance of the trained XGBoost model, as evidenced by a higher root-mean-square error (RMSE) during testing, which results in XpertAI's lower-rated explanation. 
This possibly stems from the limited size of the training dataset (only 63 data points for training). As a result, the XAI analysis reveals the model could have learned non-causal correlations within the dataset. 
Therefore, the essential features may not be faithfully represented, thereby leading XpertAI to generate a flawed explanation.

\section{Conclusion \& Outlook}
XAI is becoming increasingly important in ML workflows due to developmental, scientific, and regulatory needs. With the recent growth of XAI, more tools and techniques are emerging. In this context, we addressed a key challenge in applying XAI to chemistry - the lack of interpretability of explanations generated by XAI tools. Generally, XAI tools target technical experts, reducing usability.  
We proposed ``XpertAI'', a framework leveraging XAI and LLMs to generate intelligent natural language explanations of structure-property relationships from raw chemistry data. 
XpertAI produces readily interpretable and specific explanations while uncovering structure-property relationships.
XpertAI works for both regression and classification tasks. We showed combining XAI and LLMs is more powerful than using either alone and demonstrated that XpertAI can accurately explain input-output relationships, not just model predictions.

Importantly, XpertAI's performance depends on the surrogate model used by the XAI methods. If the surrogate model has acquired spurious data relationships, it will inevitably yield an inaccurate explanation.
In the current version of XpertAI, hyperparameters are hardcoded to enhahnce non-expert usability. 
However, technically oriented users can implement XpertAI locally using our GitHub repository.

We aim to incorporate other ML models and enable user-provided models, giving more flexibility and making XpertAI model-agnostic.
Upcoming work will integrate automated hyperparameter optimization. This will further increase accessibility for non-technical users to generate accurate explanations. 
Moreover, future work will incorporate additional open-source XAI methods and LLMs into a fully open-source XpertAI without GPT4 dependencies.

Despite current limitations, XpertAI demonstrates promising potential as an interpretable tool for uncovering novel structure-property relationships and generating scientific insights in chemistry. By leveraging AI’s strengths in explanation and language, XpertAI aims to democratize chemical knowledge extraction to accelerate scientific progress. With ongoing improvements to the framework, XpertAI represents an exciting advancement in elucidating meaningful chemical structure-property relationships to propel discovery.


\section{Code and Data Availability}
Code to XpertAI can be found at: \url{https://github.com/geemi725/XpertAI} and the XpertAI App can be found at: \url{https://xpert-ai.streamlit.app/}

\section{Acknowledgments}
We thank the expert evaluators for their contributions to this work. GPW  acknowledges funding from the EPFL large-scale Solutions4Sustainability demonstrator project (SusEcoCCUS).
PS acknowledges support from the NCCR Catalysis (grant number 180544), a National Centre of Competence in Research funded by the Swiss National Science Foundation.

\clearpage
\medskip

{
\small
\bibliography{bibliography}
}

\clearpage
\section*{Appendix A: Textual Structure-function relationships}

\subsection*{Case study 1: Presence of open metal sites in MOFs}

\subsection*{XpertAI Explanation}

\begin{itemize}
    \item The features identified by the XAI analysis affecting the Presence of open metal sites in MOFs are Fraction of Metals, Density of Solid, and Average Cationic Radius. Additional features that may be correlated with the Presence of open metal sites in MOFs include the fraction of electrons in d orbitals, the average ionization energy, and the average electron affinity (Guo et al., Not mentioned).
    \item The Fraction of Metals feature indicates the proportion of metal atoms in the MOF structure. A higher fraction of metals may increase the likelihood of open metal sites, as metals often form the nodes of MOFs where open metal sites can occur. The Density of Solid feature refers to the mass per unit volume of the MOF. A higher density may suggest a more tightly packed structure, potentially reducing the likelihood of open metal sites. The Average Cationic Radius feature represents the average size of the metal cations in the MOF. Larger cations may lead to larger pores and potentially more open metal sites (Guo et al., Not mentioned).

    \item The Presence of open metal sites in MOFs can be affected by these features in various ways. For instance, increasing the Fraction of Metals could potentially increase the number of open metal sites, assuming the additional metals form part of the MOF nodes. Similarly, decreasing the Density of Solid could potentially increase the number of open metal sites, as a less densely packed structure may have more available space for open metal sites. Increasing the Average Cationic Radius could also potentially increase the number of open metal sites, as larger cations may lead to larger pores and more open metal sites (Guo et al., Not mentioned).

    \item  In summary, the Presence of open metal sites in MOFs is influenced by various structural and chemical features of the MOFs, including the <Fraction of Metals>, Density of Solid, and Average Cationic Radius. These features can affect the likelihood of open metal sites by influencing the structure and composition of the MOFs. Understanding these relationships can help in the rational design of MOFs with desired properties (Guo et al., Not mentioned).

    \item References:
    \begin{itemize}
        \item Guo, H., Ma, J., Zhong, Y., Yin, Y., \& Chen, M. (Not mentioned). Machine learning prediction of open metal sites in metal-organic framework catalysts.
    \end{itemize}
    
\end{itemize}

\subsection*{ChatGPT Explanation}
Metal-organic frameworks (MOFs) are a class of compounds consisting of metal ions or clusters coordinated to organic ligands to form one-, two-, or three-dimensional structures. The presence of open metal sites (OMS) in MOFs is a crucial factor that determines their properties and applications, especially in gas storage, separation, and catalysis.

The presence of OMS in MOFs can be altered by several factors:

\begin{itemize}
\item Choice of Metal Ions: The type of metal ions used in the synthesis of MOFs can influence the presence of OMS. Some metal ions have a higher tendency to form OMS due to their coordination preferences. For example, MOFs made with transition metals often have OMS because these metals typically have a higher coordination number.
\item Choice of Organic Ligands: The type and structure of the organic ligands used can also affect the presence of OMS. Ligands with larger sizes or more complex structures can create more space in the MOF structure, leading to the formation of OMS. Additionally, ligands that can form strong bonds with metal ions can stabilize the OMS.
\item Synthetic Conditions: The conditions under which the MOFs are synthesized, such as temperature, pressure, and solvent, can also influence the formation of OMS. For example, lower synthesis temperatures may favor the formation of OMS because they slow down the reaction rate, allowing for more control over the MOF structure.
\item Post-Synthetic Modifications: After the MOFs are synthesized, their structures can be further modified to create or remove OMS. This can be done through methods such as solvent exchange, thermal treatment, or chemical modification.
\item Use of Modulators: Modulators are additives used during the synthesis of MOFs to control their growth and structure. By choosing appropriate modulators, the presence of OMS can be controlled.
\end{itemize}

In summary, the presence of open metal sites in metal-organic frameworks can be altered by manipulating the choice of metal ions and organic ligands, the synthetic conditions, post-synthetic modifications, and the use of modulators. These factors can be adjusted to tailor the properties of MOFs for specific applications.

\subsection*{Case study 2: Pore-limiting diameter in MOFs}

\subsection*{XpertAI Explanation }
\begin{itemize}
    \item The features identified by the XAI analysis that affect the Pore-limiting diameter in MOFs are Volume Per Atom, symmetry function G, and Unoccupied Energy Levels at Conduction Band Minimum. Other features that may be correlated with the Pore-limiting diameter in MOFs include the type of metal center, the type of ligand, and the topology of the MOF structure.
    \item The Volume Per Atom can affect the Pore-limiting diameter in MOFs as it determines the size of the atoms in the MOF structure. Larger atoms may result in smaller pore sizes (Haldoupis, Nair, \& Sholl, 2010). The symmetry function G is a measure of the symmetry of the MOF structure, which can also influence the Pore-limiting diameter. MOFs with higher symmetry may have more uniform pore sizes (Yang et al., 2022). The Unoccupied Energy Levels at Conduction Band Minimum can affect the electronic properties of the MOF, which may indirectly influence the Pore-limiting diameter.
    
    \item The Pore-limiting diameter in MOFs can be altered by changing these features. For example, using smaller atoms or increasing the symmetry of the MOF structure can increase the Pore-limiting diameter. Similarly, altering the electronic properties of the MOF by changing the Unoccupied Energy Levels at Conduction Band Minimum may also affect the Pore-limiting diameter.
    
    \item  In summary, the Pore-limiting diameter in MOFs is influenced by several features including the Volume Per Atom, symmetry function G, and Unoccupied Energy Levels at Conduction Band Minimum. These features can be manipulated to alter the Pore-limiting diameter, which can have significant implications for the performance of MOFs in various applications.
    \item References:
    \begin{itemize}
        \item Haldoupis, E., Nair, S., \& Sholl, D. S. (2010). Efficient calculation of diffusion limitations in metal organic framework materials: A tool for identifying materials for kinetic separations. Journal of the American Chemical Society, 132(21), 7528-7539.
        \item Yang, P., Lu, G., Yang, Q., Liu, L., Lai, X., \& Yu, D. (2022). Machine learning prediction of acetylene adsorption capacity in metal-organic frameworks. Journal of Physical Chemistry C, 126(4), 2159-2169.
    \end{itemize}
\end{itemize}

\subsection*{ChatGPT Explanation}
\begin{itemize}
    \item Metal-Organic Frameworks (MOFs) are porous materials that are constructed from metal ions or clusters and organic linkers. The pore limiting diameter in MOFs refers to the smallest diameter of the pores within the framework. This property is crucial as it determines the type and size of molecules that can be adsorbed or transported through the MOF.
    
    \item The pore limiting diameter in MOFs can be altered by changing the size and shape of the organic linkers and the metal ions or clusters. The organic linkers act as spacers between the metal ions or clusters, and their length, flexibility, and geometry can significantly influence the pore size. For instance, using longer linkers or linkers with larger functional groups can increase the pore size. Similarly, the choice of metal ions or clusters can also affect the pore size. Larger metal ions or clusters can lead to larger pores.
    
    \item In addition, the pore limiting diameter can also be tuned by varying the synthesis conditions, such as temperature, pressure, and solvent. These conditions can influence the assembly of the metal ions or clusters and the organic linkers, thereby affecting the pore size.
    
    \item Furthermore, post-synthetic modifications, such as the insertion of additional organic or inorganic species into the pores, can also be used to adjust the pore limiting diameter. These modifications can either increase or decrease the pore size, depending on the size and shape of the inserted species.
    
    \item In summary, the pore limiting diameter in MOFs is a tunable property that can be controlled by carefully selecting the organic linkers and metal ions or clusters, adjusting the synthesis conditions, and performing post-synthetic modifications.
\end{itemize}

\subsection*{Case study 3: Toxicity of small drug-like molecules}

\subsection*{XpertAI Explanation}

\begin{itemize}
    \item The features identified by the XAI analysis that affect the toxicity of small molecules include the presence of a heteroatom bonded to three Oxygen atoms, the presence of an aminomethyl group, and the presence of a Tertiary Amine. Additional features that may be correlated with the toxicity of small molecules include the solubility of the molecule, the presence of aromatic rings, and the presence of nucleophilic N, O, and aryl containing substructures (Sharma et al., 2017; Sharma et al., 2023).
    
    \item The presence of a heteroatom bonded to three Oxygen atoms can increase the reactivity of the molecule, potentially leading to increased toxicity. The aminomethyl group and the Tertiary Amine can also increase the reactivity of the molecule, potentially leading to increased toxicity (Kalgutkar et al., 2005). The solubility of the molecule can affect its absorption, with molecules with poor solubility showing limited absorption and hence reduced toxicity (Sharma et al., 2017). The presence of aromatic rings and nucleophilic N, O, and aryl containing substructures can also influence the toxicity of the molecule (Sharma et al., 2023).

    \item The toxicity of small molecules can be altered by changing these features. For example, the toxicity of a molecule can be reduced by decreasing the number of Oxygen atoms bonded to a heteroatom, by removing the aminomethyl group or the Tertiary Amine, or by reducing the solubility of the molecule (Sharma et al., 2017; Kalgutkar et al., 2005).
    
    \item In summary, the toxicity of small molecules is influenced by several features, including the presence of a heteroatom bonded to three Oxygen atoms, the presence of an aminomethyl group, the presence of a Tertiary Amine, the solubility of the molecule, the presence of aromatic rings, and the presence of nucleophilic N, O, and aryl containing substructures. Changes in these features can alter the toxicity of the molecule.
    
    \item  References:
    \begin{itemize}
        \item Sharma, B., Chenthamarakshan, V., Dhurandhar, A., Pereira, S., Hendler, J. A., Dordick, J. S., \& Das, P. (2023). Deep learning framework for toxicity prediction.
        \item Sharma, A. K., Srivastava, G. N., Roy, A., \& Sharma, V. K. (2017). Toxicity prediction tool for small molecules. Frontiers in Pharmacology, 8, Article 880.
        \item Kalgutkar, A. S., Shaffer, C. L., Henne, K. R., \& Mutlib, A. (2005). Bioactivation pathways of functional groups/structural motifs used in drug design. Current Drug Metabolism, 6(3), 161-185.
    \end{itemize}
\end{itemize}

\subsection*{ChatGPT Explanation}
Toxicity in small molecules is determined by their chemical structure and properties. These properties include size, shape, charge, polarity, and the presence of specific functional groups. Altering these properties can change the toxicity of the molecule.
\begin{itemize}
\item Size and Shape: The size and shape of a molecule can affect its ability to interact with biological systems. Larger molecules may have difficulty penetrating cell membranes, reducing their toxicity. Conversely, smaller molecules may easily enter cells and interact with critical components, increasing their toxicity. Changing the size or shape of a molecule can therefore alter its toxicity.
\item Charge: The charge of a molecule can also affect its toxicity. Positively charged molecules can interact with negatively charged cell membranes, potentially disrupting their function and increasing toxicity. Conversely, negatively charged molecules may be repelled by the cell membrane, reducing their toxicity. Altering the charge of a molecule can therefore change its toxicity.
\item Polarity: The polarity of a molecule can affect its solubility in water and other solvents. Polar molecules are more soluble in water, which can increase their bioavailability and therefore their potential toxicity. Non-polar molecules are less soluble in water, which can reduce their bioavailability and therefore their toxicity. Changing the polarity of a molecule can therefore alter its toxicity.
\item Functional Groups: The presence of specific functional groups can also affect a molecule's toxicity. For example, molecules with hydroxyl (-OH) groups can form hydrogen bonds with biological molecules, potentially disrupting their function and increasing toxicity. Conversely, molecules with methyl (-CH3) groups are less likely to form such bonds, reducing their toxicity. Altering the functional groups present in a molecule can therefore change its toxicity.
\item Metabolic Stability: The metabolic stability of a molecule can also influence its toxicity. Molecules that are rapidly metabolized and excreted from the body are generally less toxic, while those that persist in the body can accumulate and cause toxicity. Changing the metabolic stability of a molecule can therefore alter its toxicity.
\end{itemize}
In conclusion, the toxicity of small molecules can be altered by changing their size, shape, charge, polarity, functional groups, and metabolic stability. These changes can be made through chemical synthesis or modification.

\subsection*{Case study 4: Solubility of small molecules}
\subsection*{XpertAI Explanation }

\begin{itemize}
    \item The solubility of small molecules is a complex property that is influenced by a variety of factors. The XAI analysis has identified three key features that significantly impact the solubility of small molecules: the presence of an atom at an aromatic/non-aromatic boundary, the presence of two heteroatoms bonded to each other, and the presence of an atom with three heteroatom neighbors.
    
    \item Presence of an atom at an aromatic/non-aromatic boundary: This feature refers to the location of an atom at the boundary between an aromatic ring and a non-aromatic part of the molecule. Aromatic rings are typically hydrophobic, while non-aromatic parts of the molecule can be either hydrophilic or hydrophobic. Therefore, an atom at this boundary can influence the overall solubility of the molecule by affecting its polarity and ability to form hydrogen bonds with water molecules (Lee et al., 2023). Presence of two heteroatoms bonded to each other: Heteroatoms are atoms other than carbon and hydrogen. When two heteroatoms are bonded to each other, they can form polar bonds, which can increase the solubility of the molecule in water. This is because polar molecules are more likely to dissolve in polar solvents like water (Zhang et al., 2011). Presence of an atom with three heteroatom neighbors: An atom with three heteroatom neighbors is likely to be part of a polar functional group, which can increase the solubility of the molecule in water. This is because polar functional groups can form hydrogen bonds with water molecules, which facilitates dissolution (Cisneros et al., 2016).

    \item In conclusion, the solubility of small molecules is significantly influenced by the presence of an atom at an aromatic/non-aromatic boundary, the presence of two heteroatoms bonded to each other, and the presence of an atom with three heteroatom neighbors. By manipulating these features, it may be possible to alter the solubility of small molecules, which could have important implications for drug design and other applications.
    
    \item References:
    \begin{itemize}
        \item  Lee, S., Park, H., Choi, C., Kim, W., Kim, K. K., Han, Y. K., Kang, J., Kang, C. J., \& Son, Y. (2023). Multi-order graph attention network for water solubility prediction. Scientific Reports, 13, 957.

        \item Zhang, L., Zhu, H., Mathiowetz, A. M., \& Gao, H. (2011). Matched molecular pair analysis of changes in the free energy of binding, solubility, and lipophilicity in the course of lead optimization. Journal of Medicinal Chemistry, 54(13), 4325-4336. 

        \item Cisneros, J. A., Robertson, M. J., Mercado, B. Q., \& Jorgensen, W. L. (2016). Systematic study of effects of structural modifications on the aqueous solubility of drug-like molecules. Journal of Medicinal Chemistry, 59(3), 1090-1097.
    \end{itemize}
\end{itemize}

\subsection*{ChatGPT Explanation}
Solubility in small molecules can be altered by changing the molecular properties of the substance. This is based on the principle ``like dissolves like", which means that polar substances tend to dissolve in polar solvents, and nonpolar substances tend to dissolve in nonpolar solvents.
\begin{itemize}
    \item Polarity: The polarity of a molecule can greatly affect its solubility. Polar molecules have a positive and negative end, which allows them to interact with other polar molecules and dissolve in polar solvents like water. Nonpolar molecules, on the other hand, do not have a charge and therefore do not interact with polar molecules, making them insoluble in polar solvents.
    \item Size: The size of the molecule can also affect its solubility. Smaller molecules are generally more soluble than larger ones because they can more easily interact with the solvent molecules.
    \item Shape: The shape of a molecule can affect its solubility. Linear molecules, for example, can pack closely together and are therefore more soluble in nonpolar solvents. Branched or spherical molecules, on the other hand, cannot pack as closely and are therefore more soluble in polar solvents.
    \item Functional Groups: The presence of certain functional groups can also affect a molecule's solubility. For example, molecules with hydroxyl (-OH) or amino (-NH2) groups are more soluble in water because these groups can form hydrogen bonds with water molecules.
    \item Temperature and Pressure: The solubility of a substance can also be affected by changes in temperature and pressure. Generally, solubility increases with increasing temperature and decreases with increasing pressure.
\end{itemize}
By manipulating these properties, scientists can alter the solubility of a substance in small molecules. This is particularly useful in fields like pharmaceuticals, where the solubility of a drug can greatly affect its effectiveness.

\section*{Case study 5: Upper flammability limit of organic molecules}

\subsection*{XpertAI Explanation}
\begin{itemize}
    \item The features identified by the XAI analysis that affect the upper flammability limit of organic molecules are Structural Information Content Index (Neighborhood Symmetry of Order Zero), Information Content Index (Neighborhood Symmetry of Order Zero), the highest occupied molecular orbital (HOMO), and Dipole Moment. Additional features that may be correlated with the upper flammability limit of organic molecules could include molecular weight, number of carbon atoms, and the presence of specific functional groups.
    
    \item The Structural Information Content Index (Neighborhood Symmetry of Order Zero) and Information Content Index (Neighborhood Symmetry of Order Zero) are measures of the complexity and symmetry of the molecule's structure. Molecules with higher complexity and less symmetry may have higher upper flammability limits due to the increased number of potential reaction pathways (Yuan et al., 2019).  The highest occupied molecular orbital (HOMO) represents the energy level of the highest energy electron in the molecule. Molecules with higher HOMO energy levels may have higher upper flammability limits because they have more energy available for combustion reactions (Yuan et al., 2019). The Dipole Moment is a measure of the molecule's polarity. Molecules with higher dipole moments may have lower upper flammability limits because the increased polarity can inhibit the dispersion of the molecule in the air, reducing the likelihood of combustion (Yuan et al., 2019).
    
    \item By altering these features, such as by adding functional groups to increase the complexity of the molecule or by changing the molecule's structure to alter its dipole moment, it may be possible to influence the upper flammability limit of the molecule.

    \item In summary, the upper flammability limit of organic molecules is influenced by a variety of features, including the complexity and symmetry of the molecule's structure, the energy level of the highest energy electron, and the molecule's polarity. By understanding these relationships, it may be possible to predict and potentially control the flammability of organic compounds.

    \item References:
    \begin{itemize}
        \item Yuan, S., Jiao, Z., Quddus, N., Kwon, J. S., \& Mashuga, C. V. (2019). Developing Quantitative Structure-Property Relationship Models To Predict the Upper Flammability Limit Using Machine Learning. Industrial \& Engineering Chemistry Research, 58(12), 4905-4913.
       
    \end{itemize}
\end{itemize}

\subsection*{ChatGPT Explanation}

The upper flammability limit (UFL) of an organic molecule is the highest concentration (in terms of percentage volume in air) at which the molecule can ignite or explode. It is a critical property for safety considerations in various industries, especially those dealing with volatile organic compounds. The UFL can be influenced by several molecular properties, including molecular structure, functional groups, and molecular weight.

\begin{itemize}
    \item Molecular Structure: The structure of a molecule can significantly affect its flammability. For instance, molecules with a high degree of branching tend to have lower flammability limits due to the increased steric hindrance that prevents effective collisions with oxygen molecules. Similarly, cyclic or aromatic structures tend to have higher flammability limits due to their stability and the ease with which they can undergo combustion reactions.
    \item  Functional Groups: The presence and type of functional groups in a molecule can also influence its flammability. For example, molecules with oxygen-containing functional groups (like alcohols, ethers, and esters) generally have lower flammability limits due to the presence of oxygen, which can facilitate combustion. On the other hand, molecules with halogen-containing functional groups (like chlorides, bromides, and iodides) usually have higher flammability limits because halogens act as flame retardants.
    \item  Molecular Weight: The molecular weight of an organic molecule can also affect its flammability. Generally, as the molecular weight increases, the flammability limit decreases. This is because heavier molecules have lower vapor pressures and require higher temperatures to achieve the necessary vapor concentration for ignition.

    \item Other Factors: Other factors like temperature, pressure, and the presence of inert gases can also alter the upper flammability limit. Higher temperatures and pressures generally increase the flammability limit, while the presence of inert gases can decrease it by diluting the fuel-air mixture.

\end{itemize}
In conclusion, by altering these molecular properties, one can influence the upper flammability limit of organic molecules. This understanding is crucial in designing safer chemical processes and materials.
\clearpage
\section*{Appendix B: Claude Evaluations}

Please note that we used the question prompt given below to get ratings from \url{https://claude.ai/}. Explanation A: from XpertAI and Explanation B: from ChatGPT.

\subsection*{Question prompt:}
Compare Explanations A and B and give a score based on its interpretability and accuracy. Describe the reasons for your score.

\subsection*{Case study 1: Presence of open metal sites in MOFs}

Based on my analysis, I would rate Explanation A as more accurate and interpretable compared to Explanation B.

Reasons for my assessment:
\begin{itemize}
 \item  Explanation A provides more details on the specific features that impact the presence of open metal sites in MOFs. It lists out the top features identified by XAI analysis ('Metals Fraction', 'Density of Solid', 'Average Cationic Radius') and also mentions other correlated features ('Surface Area', 'Pore Volume', 'Functional Group'). In contrast, Explanation B only broadly refers to choice of metal ions, organic ligands, synthetic conditions etc.

\item Explanation A clearly explains the reasons behind why each feature affects open metal sites in MOFs. For example, it states that a higher metals fraction leads to more open sites because the metal ions create the open sites. Explanation B does not provide these mechanistic insights.
Explanation A links the features to the composition and structure of MOFs. It explains how manipulating the features can alter the number and distribution of open metal sites by influencing the MOF structure. Explanation B lacks these connections.
\item Explanation A provides specific recommendations for further study with relevant references. This demonstrates interpretability by enabling further research. Explanation B does not have these recommendations.
\item The language and terminology used in Explanation A is more accessible and easier to understand for a non-expert compared to Explanation B.
\end{itemize}

In summary, I would give Explanation A a higher score for accuracy and interpretability due to its detailed feature analysis, clear mechanistic explanations, structural insights, recommendations for further research, and accessible terminology. Explanation B covers the topic more broadly but lacks in-depth analysis and connections

\subsection*{Case study 2: Pore limiting diameter in MOFs}
Based on my analysis, I would rate Explanation A as more accurate and interpretable compared to Explanation B. Here are the key reasons:
\begin{itemize}
    \item Explanation A provides more specific details on the features that impact pore limiting diameter, including both the top features from XAI analysis and other correlated features. Explanation B only broadly refers to organic linkers, metal ions, and synthesis conditions.
    \item Explanation A clearly explains the quantitative relationships between each feature and pore limiting diameter (directly proportional, inversely proportional). This level of mechanistic detail is not present in Explanation B.
    \item Explanation A breaks down exactly how each feature impacts pore limiting diameter through their effects on MOF structure and properties. Explanation B lacks this level of structural analysis.
    \item Explanation A states how pore limiting diameter can be altered by changing each feature in the desired direction. Explanation B only generally mentions that the features can be changed but not how.
    \item Explanation A provides specific recommendations for further reading. This enhances interpretability. Explanation B does not have any recommendations.
    \item The terminology used in Explanation A is more accessible to a non-expert audience compared to Explanation B.
\end{itemize}
In summary, Explanation A demonstrates greater accuracy through its quantitative relationships, mechanistic insights, and specific recommendations. It also shows higher interpretability through its detailed feature analysis, structural explanations, and clear language. I would thus give Explanation A a higher score compared to Explanation B in both accuracy and interpretability.

\subsection*{Case study 3: Toxicity of small molecules}
Based on my analysis, I would rate Explanation A as more accurate and interpretable compared to Explanation B. 

Here are the key reasons:
\begin{itemize}
    \item Explanation A identifies specific features from XAI analysis and literature that impact toxicity, while Explanation B only broadly refers to size, shape, charge, etc.
    \item Explanation A provides mechanistic explanations for how each feature impacts toxicity through reactivity and formation of reactive species. Explanation B lacks this level of mechanistic detail.
    \item Explanation A states how toxicity can be altered by adding/removing specific features. Explanation B only generally states toxicity can be changed by modifying features.
    \item Explanation A uses more accessible language and defines key terms like "heteroatom" and "aminomethyl group". Explanation B uses more technical vocabulary without definitions.
    \item Explanation A provides references for further study, enhancing interpretability. Explanation B does not include any references.
    \item Explanation A demonstrates an understanding of the domain by linking features to reactivity and cellular damage. Explanation B lacks these connections to the biological effects of toxicity.
\end{itemize}
In summary, Explanation A shows greater accuracy through its use of specific features from XAI analysis, detailed mechanistic explanations, and links to biological impacts. It also demonstrates higher interpretability through accessible language, definitions, and references. I would thus give Explanation A a higher score than Explanation B in both accuracy and interpretability.

\subsection*{Case study 4: Solubility of small molecules}
Based on my analysis, I would rate Explanation A as more accurate and interpretable compared to Explanation B. 

Here are the key reasons:
\begin{itemize}
    \item Explanation A identifies specific features from XAI analysis and literature that impact solubility, while Explanation B refers broadly to polarity, size, shape, etc.
     \item Explanation A provides mechanistic explanations for how each feature affects solubility through interactions with water. Explanation B does not provide this level of mechanistic insight.
     \item Explanation A demonstrates domain knowledge by linking features to concepts like hydrophobicity, electronegativity, and hydrogen bonding. Explanation B lacks these connections.
     \item Explanation A states how solubility can be altered by manipulating specific features. Explanation B only generally states that properties can be changed.
     \item ``Explanation A uses more accessible language and defines terms like "heteroatom" and "aromatic region". Explanation B uses more technical vocabulary.
     \item Explanation A provides sources for further study, enhancing interpretability. Explanation B does not have references.
\end{itemize}
In summary, Explanation A exhibits greater accuracy through its use of specific features, detailed mechanistic explanations, and connections to chemistry concepts. It also shows higher interpretability through its accessible language, definitions, and references. I would thus give Explanation A, a higher score than Explanation B in both accuracy and interpretability.

\subsection*{Case study 5: Upper flammability of organic molecules}
Based on my analysis, I would rate Explanation B as more accurate and interpretable compared to Explanation A. 

Here are the key reasons:

\begin{itemize}
    \item Explanation B identifies specific molecular features like structure, functional groups, and molecular weight that impact upper flammability limit. Explanation A refers broadly to physicochemical properties without clear examples.
    \item Explanation B provides clear mechanistic explanations for how each feature affects flammability through combustion kinetics and thermodynamics. Explanation A lacks this level of mechanistic detail.
    \item Explanation B demonstrates strong domain knowledge by linking features to concepts like vapor pressure, steric hindrance, and flame retardants. Explanation A does not connect to these chemistry concepts.
    \item Explanation B states how upper flammability limit can be altered by modifying specific molecular features. Explanation A only generically mentions "changing these features".
    \item Explanation B uses more common terminology that is accessible to a non-expert audience. Explanation A uses highly technical vocabulary without definitions.
    \item Explanation B contextualizes the explanation in terms of safety considerations and chemical process design. Explanation A lacks this contextual framing.
\end{itemize}
In summary, Explanation B exhibits greater accuracy through its use of precise molecular features, detailed mechanistic explanations, and connections to core chemistry concepts. It also demonstrates higher interpretability through its accessible language, contextual framing, and clear statements on how to alter upper flammability limit. I would thus give Explanation B a higher score than Explanation A on both accuracy and interpretability.
\clearpage
\section*{Appendix C: Evaluation Keys }

Expert evaluators were asked to rate 2 textual explanations (from XpertAI, ChatGPT) and 1 visual explanation (SHAP plot) based on the following keys. 

\subsubsection*{Expert score:}
\begin{itemize}
    \item What can you say about the accuracy of the explanation to the best of your knowledge?
    \begin{todolist}
      \item Not accurate
      \item Somewhat accurate
      \item Accurate
      \item Cannot comment
    \end{todolist}
    \item Select all that apply
    \begin{todolist}
      \item Explanation is easy to comprehend
      \item Cannot comprehend the explanation without more information
      \item Provides scientific justifications from literature
      \item Explanation is specific to a given task/dataset
      \item Explanation is broad 
      \item Non-ML/XAI users can this explanation to describe structure-property relationships
      \item Non-ML/XAI users cannot this explanation to describe structure-property relationships
    \end{todolist}
\end{itemize}
\clearpage

\section*{Appendix D: XAI analysis plots}


\begin{figure}[h!]
\centering
\includegraphics[width=1\linewidth]{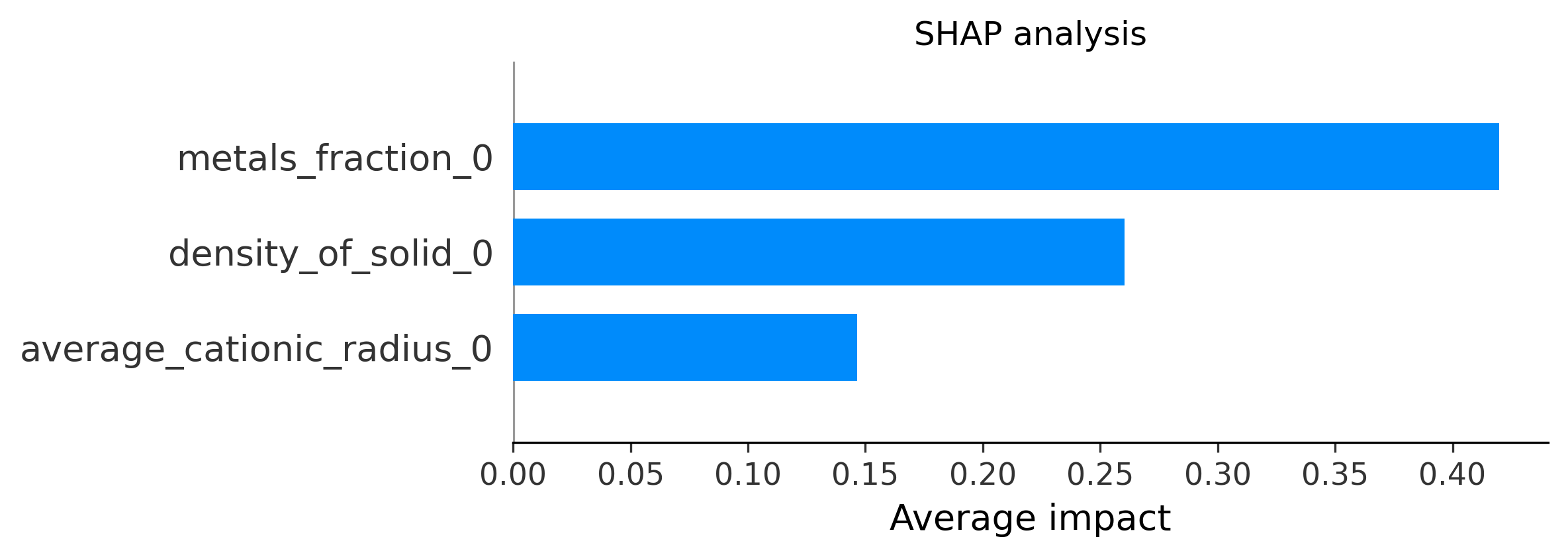}
\captionsetup{labelformat=empty}
\caption{SHAP plot for case study 1: Presence of open metal sites in MOFs}
\end{figure}


\begin{figure}[h!]
\centering
\includegraphics[width=1\linewidth]{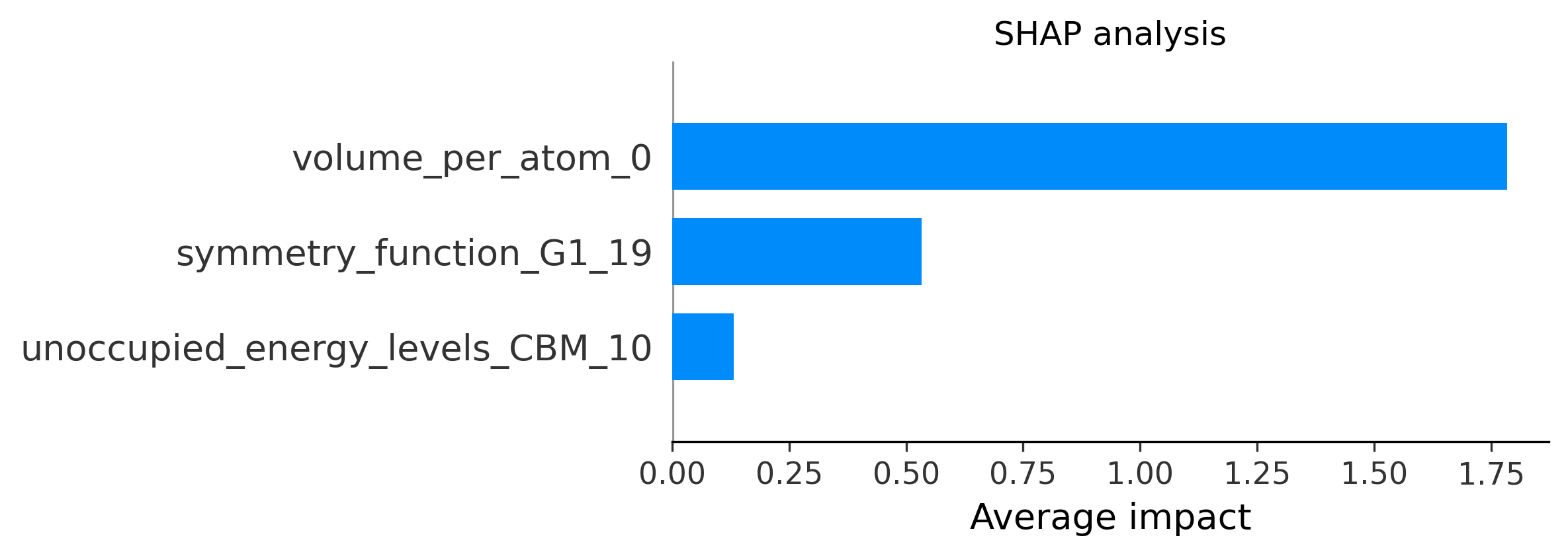}
\captionsetup{labelformat=empty}
\caption{SHAP plot for case study 2: Pore limiting diameter in MOFs}
\end{figure}

\begin{figure}[h!]
\centering
\includegraphics[width=1\linewidth]{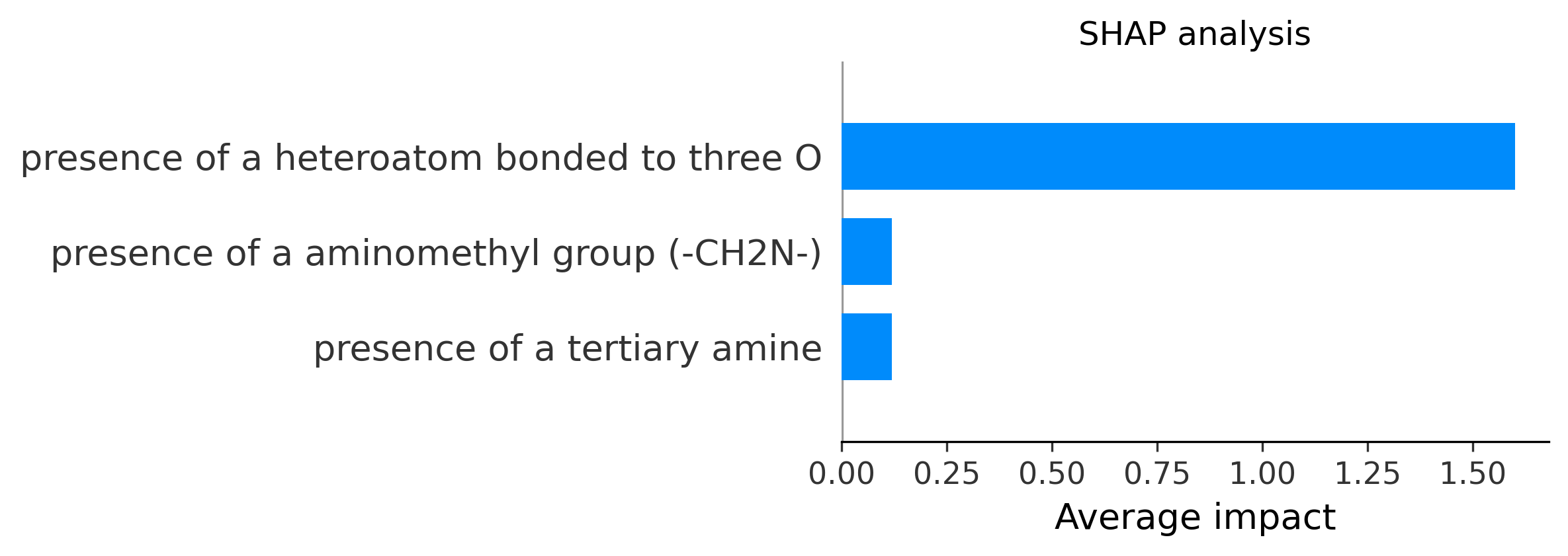}
\captionsetup{labelformat=empty}
\caption{SHAP plot for case study 3: Toxicity of small molecules}
\end{figure}


\begin{figure}[h!]
\centering
\includegraphics[width=1\linewidth]{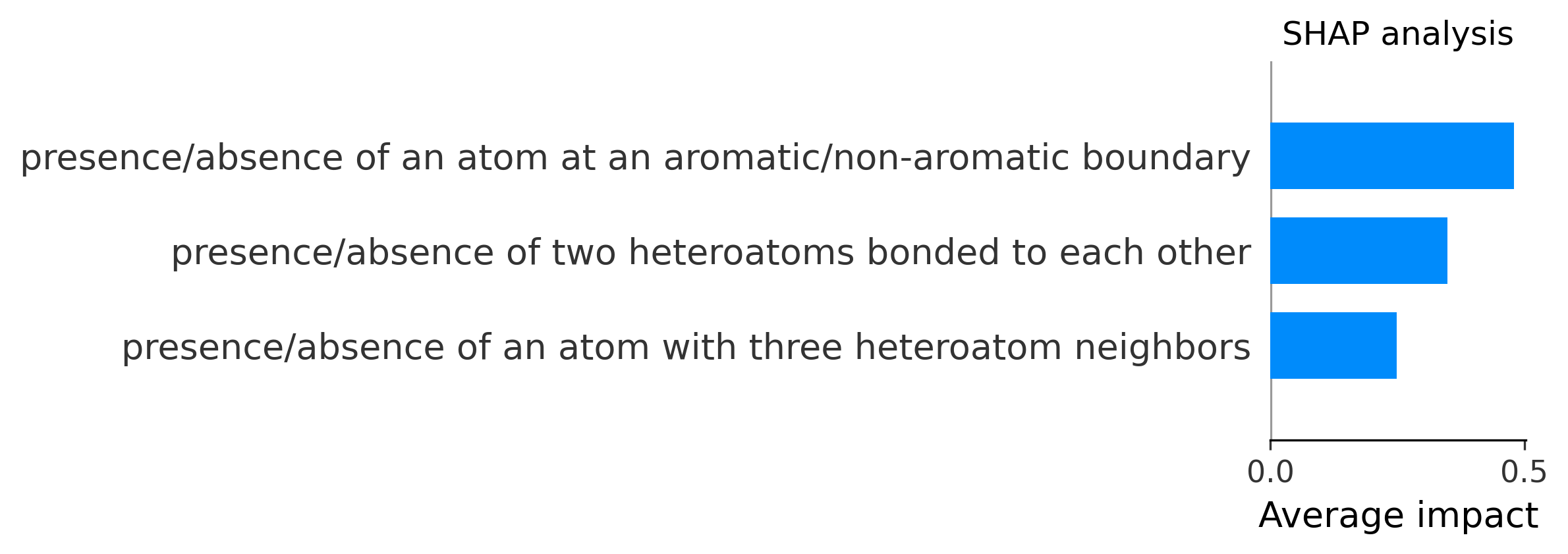}
\captionsetup{labelformat=empty}
\caption{SHAP plot for case study 4: Solubility of small molecules}
\end{figure}


\begin{figure}[h!]
\centering
\includegraphics[width=1\linewidth]{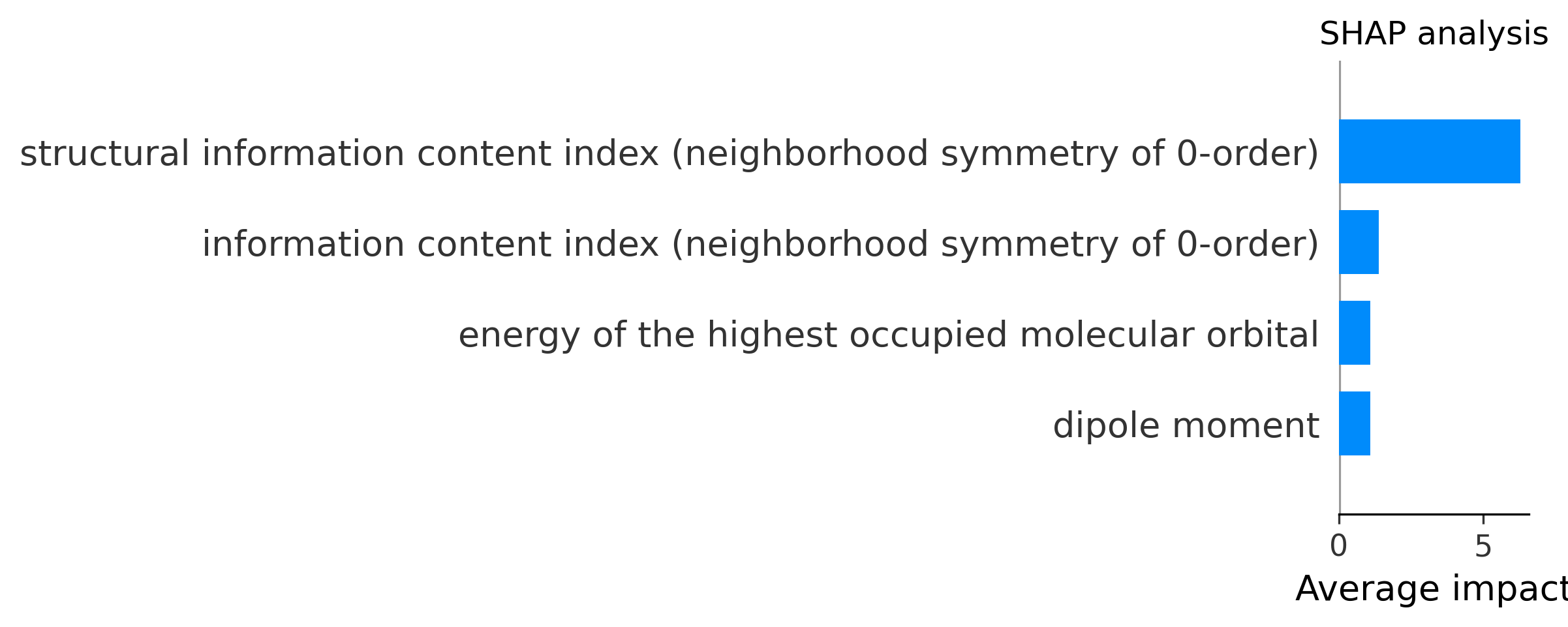}
\captionsetup{labelformat=empty}
\caption{SHAP plot for case study 5: Upper flammability of organic molecules}
\end{figure}


\end{document}